\documentclass[12pt]{article}
\usepackage[dvips]{epsfig}

\begin{document}

\author{A. de Souza Dutra$^{a,b}$\thanks{%
E-mail: dutra@feg.unesp.br}, A.C. Amaro de Faria Jr.$^{b}$ and M. Hott$^{b}$%
\thanks{%
e-mail: hott@feg.unesp.br} \\
\\
$^{a}$Abdus Salam ICTP, Strada Costiera 11, Trieste, I-34100 Italy.\\
$^{b}$UNESP-Campus de Guaratinguet\'{a}-DFQ\thanks{
Permanent Institution}\\
Departamento de F\'{\i}sica e Qu\'{\i}mica\\
12516-410 Guaratinguet\'{a} SP Brasil}
\title{{\LARGE Degenerate and critical domain walls and accelerating
universes driven by bulk particles}}
\maketitle

\begin{abstract}
We consider a scenario where our universe is taken as a three-dimensional
domain wall embedded in a five-dimensional Minkowski space-time, as
originally proposed by Brito and collaborators \cite{brito}. We explore the
existence of a richer class of soliton solutions and their consequences for
the acceleration of the universe driven by collisions of bulk particle
excitations with the walls. In particular it is shown that some of these
solutions should play a fundamental role throughout the expansion process.

\bigskip

PACS numbers: 11.27.+d, 11.10.Lm, 95.35.+d, 98.80.Cq
\end{abstract}

\newpage

\section{\protect\smallskip Introduction}

Very important systems described by quantum field theories are intrinsically
nonlinear; the Standard Model and the Quantum Chromodynamics are classical
examples. By exploring deeply such systems or effective nonlinear models,
even at the classical level, one has shown the increasing importance of the
soliton solutions (classical solutions with finite and localized energy) and
their broad applications \cite{Rajaraman}-\cite{Walgraef}. Soliton solutions
in quantum field theory describe, for instance, monopoles,\ magnetic
vortices, instantons in quantum chromodynamics, cosmic strings and magnetic
domain walls. Finding exact classical solutions, particularly solitons, is
one of the problems on nonlinear models with interacting fields. When one
has in hands a systematic method, as the one offered by Rajaraman \cite%
{rajaramanmethod}, this task becomes easier, even though one has to explore
the consequences of the classical solutions, as well as their possible
realizations in the nature. As pointed out by R. Rajaraman and E. Weinberg
\cite{rajaramanweinberg}, in such nonlinear models more than one
time-independent classical solution can exist and each one of them
corresponds to a different family of quantum states, which come into play
when one performs a perturbation around those classical solutions.

The method in \cite{rajaramanmethod}, usually called \textit{trial orbits
method}, is a very powerful one presented for finding exact soliton
solutions for nonlinear second-order differential equations of \ models with
two interacting relativistic scalar fields in 1+1 dimensions, and it is
model independent. The \textit{trial orbits method} have been applied to the
special cases whose soliton solutions of the nonlinear second-order
differential equations are equivalent to the soliton solutions of \
first-order nonlinear coupled differential equations, the so-called
Bolgomol'nyi-Prasad-Sommerfeld (BPS) topological soliton solutions\textit{\ }%
\cite{BPS}. Some years ago one of us presented a method for finding
additional soliton solutions for those special cases whose soliton solutions
are the BPS ones \cite{PLB05}. Latter, that approach was extended by
considering more general models \cite{PLB06}. Furthermore, that method shows
how to get the general equation of the orbits, explains how the different
solutions connect the different vacua of the model under analysis and, as a
novelty, presents a class of soliton solutions with a kink-like profile for
both fields with its minimum energy (BPS energy) smaller than that of the
usual solution which exhibits a kink-like configuration for one of the
fields and a lump-like configuration for the other one. Moreover, the
stability of the quantum states corresponding to these new soliton solutions
can be shown on the same basis presented in \cite{bazeia2}.

The BPS soliton solutions have found applications in a great variety of
natural systems whose dynamics can be approximately described by nonlinear
quantum field models for interacting scalar fields in 1+1 dimensions \cite%
{bazeiareview}. Those kind of models have been generalized by including into
the Lagrangian density some minimal terms that break Lorentz and CPT \
symmetries \cite{Barreto}. By analyzing the consequences of additional
soliton solutions in a given nonlinear model, some of us have shown \cite%
{lorentzbr}, by using the method developed in \cite{PLB05}, that those
nonlinear Lorentz breaking models in 1+1 dimensions exhibit additional
soliton solutions whose BPS energies are smaller than those found in \cite%
{Barreto}, and that more general Lorentz breaking models in 1+1 dimensions,
which admits soliton solutions, can be built.

In this work, we explore more deeply the classical solutions found in \cite%
{PLB05}, particularly a class of degenerate soliton solutions \cite%
{shiffman, prd2008}, for the nonlinear model of two interacting scalar
fields in 1+1 dimensions \cite{bazeiareview}. In special, we analyze the
consequences that those additional soliton solutions bring for the scenario
of accelerating universes. This scenario was recently conceived by Brito and
collaborators \cite{brito} and it is within the context of the
extra-dimensions \cite{prd2008}-\cite{bogdanos}. Our analysis is done by
following a similar approach to that of the reference \cite{brito}. It is
shown that, for a critical value of the degeneracy parameter, the reflection
coefficient of the bulk particles over the wall is maximized. That implies
into a degree of expansion of the universe which is compatible with the
present observed acceleration.

In the second section we present the model we are going to work with and
discuss the influence of each set of soliton solutions, in the context of
accelerating universes driven by the quantum states (bulk particles). In the
third section we present final comments about the impact of those new
solutions over the expanding process.

\section{Domain walls and accelerating universes}

A very important and intriguing modern physical problem is that of finding a
way to explain the observed accelerated expansion of the universe. On the
other hand, the recent cosmological data indicates that a relevant part of
the energy of the universe would be a kind of \textit{dark energy} \cite%
{darkdata}. In fact, that \textit{dark energy} is supposed to be the
responsible for that acceleration. As a consequence, many authors
look for a deep understanding of these subjects. A very interesting
possibility is the one associated to the so-called brane worlds
\cite{bogdanos}. In a recent work, Brito and collaborators
\cite{brito} conceived a scenario where our universe is taken as a
three-dimensional domain wall embedded in a five-dimensional
Minkowsky space-time, where the elastic collisions of bulk particles
with the walls would be the ultimate reason for the universe
acceleration.

The model under analysis here is precisely the same one considered in the
reference \cite{brito}, which is the scalar sector of a five-dimensional
supergravity theory obtained by means of dimensional compactification of a
higher dimensional supergravity, where the Lagrangian density \cite{brito}
looks like

\begin{equation}
\mathcal{L}=| det g_{\mu \nu}|^{\frac{1}{2}}\left\{ -\frac{1}{4}M^3R_{5}+
G_{AB}\left[ \partial_{\mu}\phi^{A}\partial^{\mu}\phi^{B} -\frac{1}{4} \frac{%
\partial W(\phi )}{\partial \phi^{A}}\frac{\partial W(\phi )}{\partial
\phi^{B}}\right] +\frac{1}{3\, M^3}W(\phi )^2\right\} ,
\end{equation}

\noindent where $1/M$ is the five-dimensional Planck length, $R_{5}$ is the
Ricci scalar, and $G_{AB}$ is the metric on the scalar target space. Now,
considering the limiting situation where $M>>1<<M_{Pl}$, so that the
five-dimensional gravity decouples from the scalars, one is left with a
3-brane which is a classical $3d$ domain wall embedded in a $5d$ Minkowski
space \cite{brito}. Thus, in this limit, and choosing the case with two
scalar fields in the target space, we are working with a nonlinear model
with interacting scalar fields in 1+1 dimensions whose time-independent
equations of motion (Eqs. (\ref{2ndorder}) below), are those one-dimensional
static equations for the scalar fields taken into account in the reference
\cite{brito}. We follow the same route as in that reference, but here we
consider more general set of static solutions and compare our results with
those obtained in \cite{brito} where the bulk particles, which are the
quantum excitations around the classical solutions, collide with a 3d domain
wall described by a kink. The essential idea is to show that the situation
is richer than analyzed in \cite{brito}, and that from a complete solution
as the one we discuss here, important consequences for the expanding
scenario show up.

The model was introduced before to study BPS soliton solutions \cite{bazeia1}
and, recently, it was explored in the context of brane worlds \cite{prd2008}%
. It consists of two interacting real scalar fields in 1+1 dimensions and it
is described by the Lagrangian density
\begin{equation}
\mathcal{L}=\frac{1}{2}(\partial _{\mu }\phi )^{2}+\frac{1}{2}(\partial
_{\mu }\chi )^{2}-V(\phi ,\chi ),  \label{calL}
\end{equation}%
where the potential is given by
\begin{equation}
V(\phi ,\chi )=\frac{1}{2}\lambda ^{2}(\phi ^{2}-a^{2})^{2}+(2\mu
^{2}+\lambda ~\mu )\phi ^{2}\chi ^{2}-\lambda ~\mu ~a^{2}~\chi ^{2}+\frac{1}{%
2}\mu ^{2}\chi ^{4}.  \label{potential}
\end{equation}%
The distinctive property of this model is that its potential can be written
in terms of a so-called superpotential as
\begin{equation}
V(\phi ,\chi )=\frac{1}{2}\left( \frac{\partial W(\phi ,\chi )}{\partial
\phi }\right) ^{2}+\frac{1}{2}\left( \frac{\partial W(\phi ,\chi )}{\partial
\chi }\right) ^{2},  \label{potential2}
\end{equation}%
where the superpotential is%
\begin{equation}
W(\phi ,\chi )=\phi \left[ \lambda \left( \frac{\phi ^{2}}{3}-a^{2}\right)
+\mu \chi ^{2}\right] .  \label{superpot}
\end{equation}%
Hence, finding the classical solutions with minimum energy for the
time-independent equations of motion
\begin{equation}
\frac{d^{2}\phi }{dx^{2}}=\frac{\partial V}{\partial \phi }\hspace{0.5in}%
\mathrm{and\hspace{0.5in}}\frac{d^{2}\chi }{dx^{2}}=\frac{\partial V}{%
\partial \chi },  \label{2ndorder}
\end{equation}%
is equivalent to find the classical solutions with minimum energy for the
time-independent first-order differential equations
\begin{equation}
\phi ^{\prime }=W_{\phi }(\phi ,\chi ),~\chi ^{\prime }=W_{\chi }(\phi ,\chi
).  \label{1storder1}
\end{equation}%
In the above equations the prime means the derivative with respect to the
space coordinate, and $W_{\phi }~(W_{\chi })$ stands for the partial
derivative of $W(\phi ,\chi )$ with respect to the $\phi ~(\chi )$ field.
The minimum energy (BPS energy) \cite{BPS} for nonlinear systems described
by the Lagrangian density (\ref{calL}) with potentials written as (\ref%
{potential2}) is found to be given by
\begin{equation}
E_{BPS}=|W\left( \phi _{j},\chi _{j}\right) -W\left( \phi _{i},\chi
_{i}\right) |\ ,  \label{Ebps}
\end{equation}%
where $\phi _{i}$ and $\chi _{i}$ mean the $i-th$ vacuum states of the model.

Applying the method developed in \cite{PLB05}, we note that it is possible
to write the relation $d\phi /W_{\phi }=dx=d\chi /W_{\chi }$, where the
differential element $dx$ is a kind of invariant. Thus, one is lead to

\begin{equation}
\frac{d\phi }{d\chi }=\frac{W_{\phi }}{W_{\chi }}.  \label{eqm}
\end{equation}%
This is in general a nonlinear differential equation relating the scalar
fields of the model. If one is able to solve it completely for a given
model, the function $\phi \left( \chi \right) $ can be used to eliminate one
of the fields, rendering the equations (\ref{1storder1}) uncoupled and
equivalent to a single one. Finally, the resulting uncoupled first-order
nonlinear equation can be solved in general, even if numerically. By
substituting the derivatives of the superpotential (\ref{superpot}) with
respect to the fields in (\ref{eqm}), it can be rewritten as a linear
differential equation,
\begin{equation}
\frac{d\rho }{d\chi }-\frac{\lambda }{\mu ~\chi }=\chi ,  \label{eqm3}
\end{equation}%
by the redefinition of the field, $\rho =\phi ^{2}-a^{2}$. Now, the general
solutions are easily obtained as
\begin{equation}
\rho (\chi )=\phi ^{2}-a^{2}=c_{0}~\chi ^{\lambda /\mu }-\frac{\mu }{\lambda
-2\mu }~\chi ^{2},\hspace{0.5in}\mathrm{for\ \ }\lambda \neq 2\mu ,
\label{sol1eqm}
\end{equation}%
\noindent where $c_{0}$ is an arbitrary integration constant and, for the
sake of simplicity, we will restrict our study to the case where $\lambda
\neq 2\mu $. Then, we substitute the above solutions in the differential
equation (\ref{1storder1}) and obtain the following first-order differential
equation for the field $\chi (x)$, obtaining%
\begin{equation}
\frac{d\chi }{dx}=\pm \,2\,\mu \chi \sqrt{\,a^{2}+c_{0}\,\chi ^{\lambda /\mu
}-\frac{\mu }{\lambda -2\mu }~\chi ^{2}}\,,\quad \,\,\lambda \neq 2\mu .
\label{eqdchi1}
\end{equation}

\noindent Despite the fact that in general explicit solutions for the above
equation can not be obtained, one can verify numerically that the solutions
belong to the same classes, and some of those classes of solutions can be
written in closed explicit forms. In those last cases we are able to obtain
the several types of soliton solutions, whose consequences we discuss below.

The set of solutions used by Brito \textit{et al} \cite{brito}, called here
type-A kink \cite{PLB05} to distinguish it from other types of kink
solutions we present for this model, can be obtained by means of the method
described in the previous section just by taking $c_{0}=0$ in the expression
(\ref{eqdchi1}). In this case that equation can be solved analytically for
any value of $\lambda $ and $\mu $, in the range $\lambda >2\,\mu $. We have
the following solutions for $\chi (x)$ and $\phi \left( x\right) $:%
\begin{equation}
\bar{\chi}_{A}(x)=a\sqrt{\frac{\lambda -2\mu }{\mu }}\mathrm{sech}(2\mu ax),~%
\bar{\phi}_{A}(x)=\pm a~\tanh (2\mu ax).  \label{chi2a}
\end{equation}%
In this case the BPS energy is given by $E_{BPS}^{A}=\frac{4}{3}\lambda
a^{3} $.

Other soliton solutions can be found when one considers the integration
constant $c_{0}\neq 0$ \cite{shiffman}. It was found in \cite%
{PLB05,PLB06,lorentzbr,prd2008} that in three particular cases the equation (%
\ref{eqdchi1}) can be solved analytically. For $c_{0}<-2a$ and $\lambda =\mu
$ it was found that the solutions for the $\chi (x)$ field are lump-like
solutions, which vanishes when $x\rightarrow \pm \infty $. On its turn, the
field $\phi (x)$ exhibits a kink-like profile. They also connect the vacua
of the model and also have BPS energy $E_{BPS}^{A}=\frac{4}{3}\lambda a^{3}$%
. These classical solutions can be written as
\begin{equation}
\bar{\chi}_{A}^{(1)}(x)=\frac{2a}{\sqrt{c_{0}^{2}-4}\cosh (2\mu ax)-c_{0}}~~%
\mathrm{and~}~\bar{\phi}_{A}^{(1)}(x)=a\frac{\sqrt{c_{0}^{2}-4}\sinh (2\mu
ax)}{\sqrt{c_{0}^{2}-4}\cosh (2\mu ax)-c_{0}},  \label{chi2atil1}
\end{equation}

An interesting aspect of these solutions is that, for some values of $%
c_{0}\,<-2a$, $\bar{\phi}_{A}^{(1)}(x)$ exhibits a double kink profile. We
can speak of a formation of a double wall structure. In Figure 1 we plot
some typical profiles of the soliton solutions in the case where $\lambda
=\mu $, both when $c_{0}$ is close to its critical value ($c_{0}=-2$ in this
case) and far from it. Both fields are there represented. One can verify
that the distance from one wall to the other one increases as $c_{0}$
approaches its critical value. For the critical value of $c_{0}$ the double
wall structure merges into a single one.

Similar behavior is also noted in the classical solutions for $\lambda =4\mu
$ and $c_{0}<1/16$. In this case the fields profile look like
\begin{equation}
\bar{\chi}_{A}^{(2)}(x)=-\frac{2a}{\sqrt{\sqrt{1-16c_{0}}~\cosh (4\mu ax)+1}}%
,\hspace{0.5in}  \label{chi2atil2}
\end{equation}%
and%
\begin{equation}
\bar{\phi}_{A}^{(2)}(x)=\sqrt{1-16c_{0}}a\frac{\sinh (4\mu ax)}{\sqrt{%
1-16c_{0}}~\cosh (4\mu ax)+1}.\hspace{0.5in}  \label{double2}
\end{equation}%
In the above solution one can also see the double kink profile for some
values of $c_{0}$, and the increasing of the distance from one wall to the
another as $c_{0}$ approaches its critical value ($c_{0}$ = 1/16 in this
case). Once again, at the critical value of $c_{0}$, the double wall
structure coalesces into a single wall. In Figure 2 it is shown the behavior
of the energy density for $\lambda =4\mu $. From that, it is quite evident
the appearance of the double walls when one approaches $c_{0}=1/16a^{2}$.

Finally, very interesting analytical soliton solutions were shown to exist
when one takes $\lambda =\mu $ and the critical parameter $c_{0}=-2a$ and
for $\lambda =4\mu $ and the critical parameter $c_{0}=1/16a^{2}$, in the
equation (\ref{eqdchi1}). The novelty in these cases is the fact that both,
the $\chi (x)$ field and the $\phi (x)$ field present a kink-like profile
and the BPS energy is half of that of type-A kinks, that is $E_{BPS}^{B}=%
\frac{2}{3}\lambda a^{3}$. We call this set of solutions as type-B kinks.
For $\lambda =\mu $ and $c_{0}=-2a$ the classical solution for the fields
can be shown to be given by
\begin{equation}
\bar{\chi}_{B}^{(1)}(x)=\frac{a}{2}(1\pm \tanh (\mu a\,x))~~\mathrm{and~}~%
\bar{\phi}_{B}^{(1)}(x)=\frac{a}{2}(\tanh (\mu ax)\mp 1).  \label{chi2b}
\end{equation}%
For $c_{0}=1/16a^{2}$ and $\lambda =4\mu $, the following set of type-B
kinks is obtained
\begin{equation}
\bar{\chi}_{B}^{(2)}(x)=-\sqrt{2}a\frac{\cosh (\mu ax)\pm \sinh (\mu ax)}{%
\sqrt{\cosh (2\mu ax)}}~~\mathrm{and~}~\ \bar{\phi}_{B}^{(2)}(x)=\frac{a}{2}%
(1\mp \tanh (2\mu ax)).  \label{chi2btil}
\end{equation}%
The type-B solutions also connect the vacua of the model, but this time one
could interpret these solutions as two kinds of torsion in a chain,
represented through an orthogonal set of coordinates $\phi $ and $\chi $. In
the plane ($\phi $,$\chi $), the type A kinks correspond to a complete
torsion going from $(-a,0)$ to $(a,0)$ whereas the type B corresponds to a
half torsion, where the system goes from $(-a,0)$ to $(0,a)$, in the case
where ($\lambda =\mu $) for instance.

Now, by proceeding as in \cite{brito}, we perform a linear perturbation of
the $\chi (r,t)$ field around the classical solutions, that is
\begin{equation}
\chi (r,t)=\bar{\chi}(r)+\zeta (r,t)~\mathrm{and}~\phi (r,t)=\bar{\phi}(r),
\label{fluc2}
\end{equation}%
where $\bar{\chi}(r)$ and $\bar{\phi}(r)$ are the classical solutions
(background fields) and $\zeta (r,t)$ is the quantum field. By expanding the
action up to quadratic terms in the quantum fields we obtain second-order
differential equations for the quantum fields
\begin{equation}
\partial _{\mu }\partial ^{\mu }\zeta +\bar{V}_{\chi \chi }(r)\zeta =0,
\label{eqmotion}
\end{equation}%
that is, the quantum field obeys a Klein-Gordon equation with an effective
potential $\bar{V}_{\chi \chi }(r)$ which is obtained by taking the second
derivative of the potential given in (\ref{potential}) with respect to $\chi
(r)$ and evaluated at the classical solutions $\bar{\chi}(r)$ and $\bar{\phi}%
(r)$%
\begin{equation}
\bar{V}_{\chi \chi }(x)=2(2\mu ^{2}+\lambda ~\mu )\bar{\phi}^{2}+6\mu ^{2}%
\bar{\chi}^{2}-2\lambda ~\mu ~a^{2}.  \label{effpot}
\end{equation}

If we consider our model as a five-dimensional one and that the fluctuations
can be expanded in terms of plane-waves in the space-time coordinates, that
is
\begin{equation}
\zeta (r,t)=\zeta (r)\exp [-i(\omega t-k_{x}x-k_{y}y-k_{z}z),
\label{planewave}
\end{equation}%
with $\zeta (r)$ a function of the fifth coordinate, we obtain the following
Schr\"{o}dinger-like equation
\begin{equation}
\left( -\frac{d^{2}}{dr^{2}}+\bar{V}_{\chi \chi }(r)\right) \zeta
(r)=\varepsilon ~\zeta (r),  \label{sch-eq}
\end{equation}%
where $\varepsilon =\omega ^{2}+k_{x}^{2}+k_{y}^{2}+k_{z}^{2}$ and we are
considering that the static solutions described in the previous section are
now functions of the coordinate $r$.

Now, we compare the effect of all the above soliton solutions over the
acceleration scenario proposed in \cite{brito}. For this, we begin by
considering the static solutions given by the equations (\ref{chi2atil1}).
By substituting them in the effective potential expressed in (\ref{effpot}),
we find that the Schr\"{o}dinger-like equation for the quantum excitations
is
\begin{equation}
\left( -\frac{d^{2}}{dr^{2}}+V_{eff}^{(A-1)}(r)\right) \zeta (r)=\varepsilon
~\zeta (r),  \label{sch-eq1}
\end{equation}

\noindent where
\begin{equation}
V_{eff}^{(A-1)}(r)=6\mu ^{2}a^{2}\frac{(c_{0}^{2}-4)\sinh ^{2}(2\,\mu
\,a\,r)+4}{\left( \sqrt{c_{0}^{2}-4}\cosh (2\,\mu \,a\,r)-c_{0}\right) ^{2}}%
-2\mu ^{2}a^{2},  \label{veff1}
\end{equation}

\noindent and, when $c_{0}=-2$, the effective potential, for $\mu =\lambda $%
, becomes
\begin{equation}
V_{eff}^{(B-1)}(r)=\mu ^{2}\,a^{2}(1+3\;\tanh ^{2}(\mu \,a\,r)).
\end{equation}

\noindent Note that, in the case analyzed in \cite{brito}, the effective
potential is
\begin{equation}
V_{eff}^{(A)}(r)=4\,\mu \,a^{2}-4\,\mu \,a^{2}\left( 4-\frac{\lambda }{\mu }%
\right) \mathrm{sech}^{2}\left( 2\,\mu \,a\,r\right) .  \label{eff-pot-brito}
\end{equation}

In Figure 3 it is shown the behavior of the effective potential coming from
the domain wall solutions in four situations: the case studied by Brito et
al. and for the case of degenerate solitons when $c_{0}$ is far from its
critical value, near it and at the critical point ($c_{0}=-2$). From that
Figure one can perceive that, apart from the appearance of the two wells
potential, for $c_{0}$ close to the critical value, the case where the
reflection coefficient is bigger is probably the case studied in \cite{brito}%
.

The more remarkable result comes to life for $\lambda =4\mu $. By
substituting the equations (\ref{chi2atil2}) and (\ref{double2}) in (\ref%
{effpot}) , one can verify that the quantum excitations of the $\chi (r,t)$
field satisfies the effective Schr\"{o}dinger equation (\ref{sch-eq}) with
an effective potential given by
\begin{equation}
V_{eff}^{(A-2)}(r)=12\mu ^{2}a^{2}\frac{(1-16c_{0})\sinh ^{2}(4\,\mu
\,a\,r)+2}{\left( \sqrt{1-16c_{0}}\cosh (4\,\mu \,a\,r)+1\right) }-8\mu
^{2}a^{2},  \label{veff2}
\end{equation}

\noindent and, when $c_{0}=1/16$, the effective potential becomes
\begin{equation}
V_{eff}^{(B-2)}(r)=\mu ^{2}\,a^{2}\,\mathrm{sech}^{2}(2\,\mu \,a\,r)\left[
2+5\,\cosh (4\,\mu \,a\,r)+3\,\sinh (2\,\mu \,a\,r)\right] .
\end{equation}%
It can be observed from equation \ref{eff-pot-brito} that, for $\lambda
=4\mu $, one has a constant effective potential $V_{eff}^{(IIA)}(r)=4\,\mu
\,a^{2}$, so there is no reflection of the bulk particles. This is an
artifact of the particular solution considered in\ \cite{brito}. In contrast
with this, the solutions presented here lead to effective potentials such
that the reflection coefficient increases as $c_{0}$ approaches its critical
value. Those effective potentials are presented in Figure 4. One can note
that, for $c_{0}$ far from and close to its critical value, there is a
non-zero probability for the bulk particles to be transmitted, whereas, for
the critical case, the effective potential acquires a ramp-like profile, so
granting that the bulk particles with energy below the height of the ramp
are certainly reflected by that \textquotedblleft infinitely large
barrier\textquotedblright .

For a given amount of energy available to build up structures in the
original universe, we argue that it should be distributed, for instance, like

\begin{equation}
{\mathcal{E}_{T}}=\sum_{i=1}^{M}N^{(i)}{\mathcal{E}}_{BPS}^{(i)},
\end{equation}

\noindent where $N^{(i)}$ stands for the number of defect structures of the $%
i$-th specimen and

\begin{equation}
{\mathcal{E}}_{BPS}^{(i)}=E_{BPS}^{(i)}+\frac{k_{r}^{2}}{2m},
\end{equation}

\noindent with $E_{BPS}^{(i)}$ being the BPS energy of a given soliton
configuration, and $k_{r}$ the momentum of the excitation with mass $m$
hitting the wall.

In order to estimate the force acting on the wall by the incoming
excitations, we consider quasi-elastic collisions to guarantee the energy
conservation in the process. In other words, the kinetic energy acquired by
the wall comes from the hitting particles. In that case, the momentum
transferred to the wall is given by

\begin{equation}
\Delta P\cong N_{reflected}(2\,k_{r})=N_{TI}\,R(2k_{r}),
\end{equation}

\noindent where $R$ is the reflection coefficient and $N_{TI}$ is the total
number of incoming particles. From this, we can conclude that the
acceleration of the wall is

\begin{equation}
a_{wall}=\frac{1}{M_{wall}}\frac{\Delta P}{\Delta t}=\alpha \,R,
\end{equation}

\noindent with%
\[
\alpha \cong \frac{2~k_{r}}{M_{wall}}\frac{N_{TI}}{\Delta t},
\]
and $\frac{N_{TI}}{\Delta t}$ is the flux of incident particles.

From now on we follow the brane inflation scenario presented in \cite{brito}%
, and similarly in \cite{dvali4}. In that scenario one supposes that we live
in a three-dimensional domain wall which is considered as the surface of a
hypersphere, whose radius increases due to the hitting bulk particles. The
increasing of the radius can be modeled as $r(t)=a(t)~\tilde{r}$, where $%
a(t) $ is the known scale factor and $\tilde{r}$ is the constant radius of
the bubble in the comoving system. Consequently, in this cosmological model,
we have the metric of the 4d universe given by%
\begin{equation}
ds_{4}^{2}=dt^{2}-a(t)^{2}(dx^{2}+dy^{2}+dz^{2}).
\end{equation}%
From this point of view, the acceleration of the wall is proportional to the
second-derivative in time of the scale factor, that is $a_{wall}=\ddot{a}(t)%
\tilde{r}$, and we have the following relation%
\begin{equation}
\ddot{a}(t)=\kappa \,R.
\end{equation}%
where $\kappa =\alpha /\tilde{r}$.

It can be seen from Figure 4 that the effective potential barrier width
increases when the degeneracy parameter $c_{0}$ approaches its critical
value $c_{0}^{\ast }$. As a consequence, the reflection coefficient becomes
asymptotically constant. Conversely, for $c_{0}$ far from the critical value
the reflection coefficient tends to zero, since the width and height of the
barrier diminish in this limiting situation. Thus, when $c_{0}$ is close to
the critical value, one has%
\begin{equation}
a\left( t\right) \sim \frac{1}{2}k~t^{2},
\end{equation}%
which is compatible with the present observed behavior of the universe.
Furthermore, in the opposite limit, when $c_{0}~$is far from the critical
value, the reflection coefficient approaches zero. In this situation one can
suppose that this happens exponentially, so the scale factor is given by%
\begin{equation}
a\left( t\right) \sim \frac{\beta }{2}~e^{\gamma ~t},
\end{equation}%
as requested by an inflationary era.

At this point it is important to remark that $c_{0}^{\ast }$ is a function
of the parameters $\mu $ and $\lambda $. This can be noted from the fact
that its value changes when the relation among $\lambda $ and $\mu $ changes
(see the cases analyzed above). This property allows us to suppose that $%
c_{0}^{\ast }$ would have a time dependence, as well as the parameters of
the potential. That dependence might have its origin in the universe
expansion itself. The temperature of the universe decreases along with its
expansion and, consequently, the parameters of the potential would exhibit a
temperature dependence. Therefore, $c_{0}^{\ast }$ approaches the value of $%
c_{0}$ the universe was created with.

\bigskip

\section{Final remarks}

In this work we have analyzed the impact of a general set of soliton
solutions over the reflection coefficient of the bulk particle collisions
with a 3d domain wall, as originally proposed in \cite{brito}. We have shown
that when the potential parameters are such that $\lambda =4\mu $, the
effective potential interacting with the bulk particles have its reflection
coefficient arbitrarily larger, depending on the value of the degeneracy
parameter $c_{0}$. A remarkable situation is that when $c_{0}$ reaches its
critical value. In that case, the effective potential becomes a kind of step
potential, so that the bulk particles having energy lesser than that of the
step potential will always be reflected, so producing a quadratic driven
acceleration. The degenerate structure, however, in principle would be
created with an arbitrary $c_{0}$, far from the original critical value of
the universe, so leading to an exponentially growing driven acceleration.
Therefore, while the universe is cooling due to its expansion, the critical
value $c_{0}^{\ast }(\lambda (T),\mu (T))$ approaches $c_{0}$ and, as a
consequence, the reflection coefficient increases and the acceleration also
becomes quadratic. Thus, supposing that both types of topological structures
were created in the origin of the universe, one can see that the degenerate
structure would dominate the beginning of the process. This happens due to
the fact that, when the reflection coefficient is close to one, the scale
factor $a(t)\sim t^{2}$ and for an exponentially growing reflection
coefficient, it grows exponentially too. Therefore, in the beginning the
degenerate structure dominates the expansion and, as the time goes by, both
structures predict a quadratic behavior, which is compatible with the
present experimental bounds.

\bigskip

\textbf{Acknowledgements: }The authors ASD and MH thanks to CNPq and ACAF to
CAPES for the partial financial support. We also thanks to Professor D.
Bazeia for introducing us to the matter of solitons and BPS solutions. This
work has been partially done during a visit (ASD) within the Associate
Scheme of the Abdus Salam ICTP.

\noindent

\newpage

\begin{figure}[tbp]
\begin{center}
\begin{minipage}{20\linewidth}
\epsfig{file=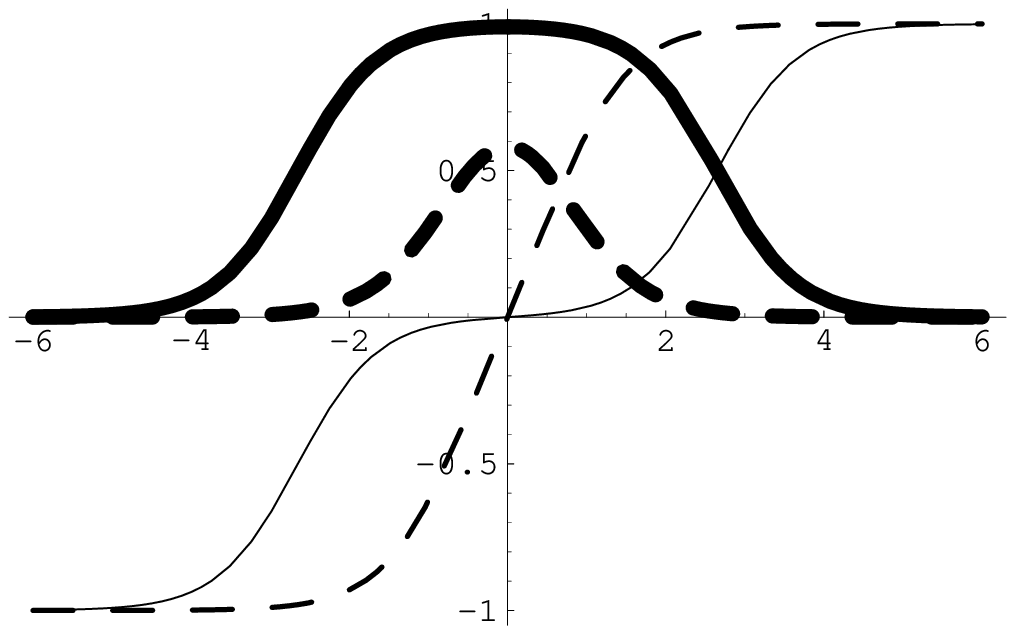}
\end{minipage}
\end{center}
\caption{Typical profiles of the soliton solutions in the case where $%
\protect\lambda = \protect\mu$ ($a=1$) as a function of $x$, both when $c_0$
is close to its critical value ($c_0=-2$ in this case) (solid lines) and far
from it (dashed lines). The $\protect\chi$ field is represented on the thick
lines and $\protect\phi$ on the thin ones.}
\label{fig:fig1}
\end{figure}

\begin{figure}[tbp]
\begin{center}
\begin{minipage}{20\linewidth}
\epsfig{file=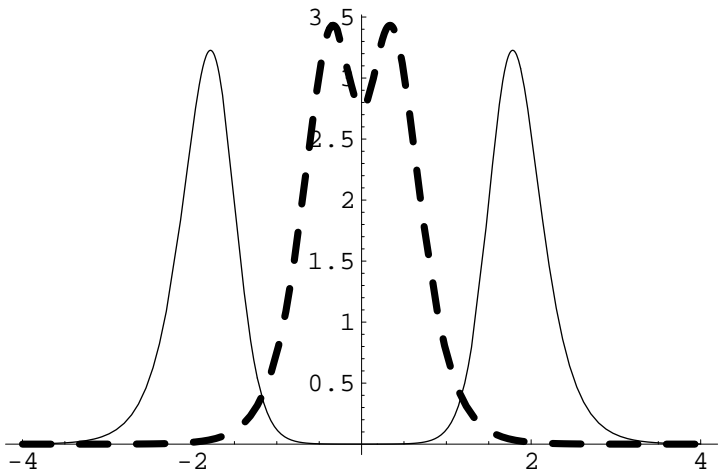}
\end{minipage}
\end{center}
\caption{Energy densities for the case where $\protect\lambda = 4\protect\mu$%
, where $c_0$ is near its critical value (solid line) ( $c_0 = 1/16$ in this
situation), and when it is far from this value (dashed line). Here it is
evident the appearance of the double wall structure.}
\label{fig:fig2}
\end{figure}

\begin{figure}[tbp]
\begin{center}
\begin{minipage}{20\linewidth}
\epsfig{file=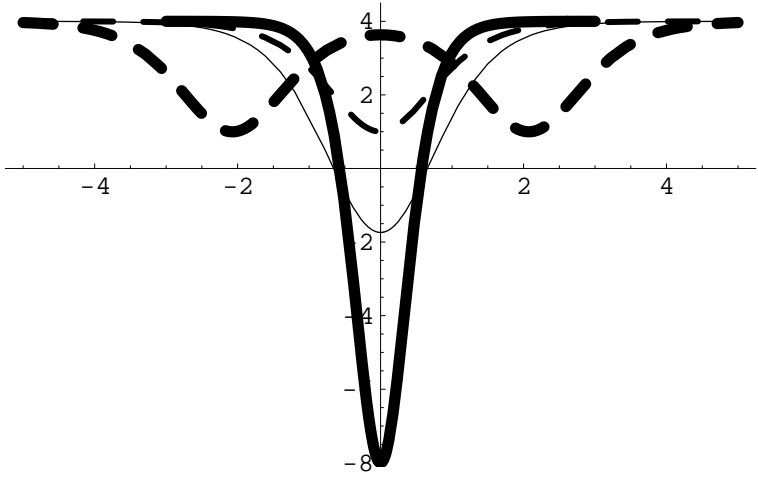}
\end{minipage}
\end{center}
\caption{Comparison of the domain wall effective potentials for the case of
Brito et al. (thick solid line) and degenerate solitons when $c_0$ is far
from its critical value (thin solid line), near it (thick dashed line) and
at the critical point ($c_0=-2$) (thin dashed line). All of them are
depicted for the case where $\protect\lambda = \protect\mu$.}
\label{fig:fig3}
\end{figure}

\begin{figure}[tbp]
\begin{center}
\begin{minipage}{20\linewidth}
\epsfig{file=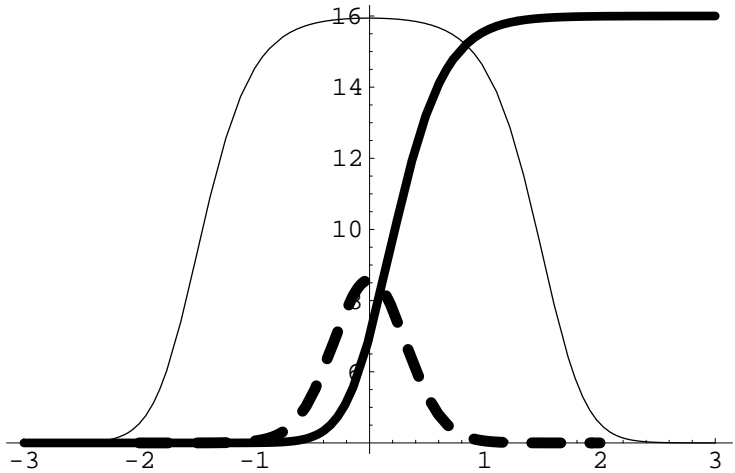}
\end{minipage}
\end{center}
\caption{Comparison of the domain wall effective potentials for the case
where $\protect\lambda = 4 \protect\mu$. The thick solid line represents the
critical case ($c_0=1/16$). The other curves correspond to $c_0$ far from
its critical value (thick dashed line) and near it (thin solid line).}
\label{fig:fig4}
\end{figure}

\end{document}